\documentclass[sigconf]{acmart}
\usepackage[nolist,nohyperlinks]{acronym}
\usepackage{multirow}
\usepackage{graphicx}
\usepackage{amsmath}
\AtBeginDocument{%
  }

\copyrightyear{2025}
\acmYear{2025}
\setcopyright{acmlicensed}\acmConference[RecSys '25]{Proceedings of the Nineteenth ACM Conference on Recommender Systems}{September 22--26, 2025}{Prague, Czech Republic}
\acmBooktitle{Proceedings of the Nineteenth ACM Conference on Recommender Systems (RecSys '25), September 22--26, 2025, Prague, Czech Republic}
\acmDOI{10.1145/3705328.3748018}
\acmISBN{979-8-4007-1364-4/2025/09}

\begin{document}


\title{``Beyond the past'': Leveraging Audio and Human Memory for Sequential Music Recommendation}

\author{Viet-Anh Tran}
\authornote{Contact author: \href{research@deezer.com}{research@deezer.com}}
\affiliation{
  \institution{Deezer Research}
  \city{}
  \country{}
}

\author{Bruno Sguerra}
\affiliation{
  \institution{Deezer Research}
  \city{}
  \country{}
}

\author{Gabriel Meseguer-Brocal}
\affiliation{
  \institution{Deezer Research}
  \city{}
  \country{}
}

\author{Lea Briand}
\affiliation{
  \institution{Deezer Research}
  \city{}
  \country{}
}

\author{Manuel Moussallam}
\affiliation{
  \institution{Deezer Research}
  \city{}
  \country{}
}

\renewcommand{\shortauthors}{Viet-Anh Tran, Bruno Sguerra, Gabriel Meseguer-Brocal, Lea Briand \& Manuel Moussallam}

\begin{acronym}
    \acro{DL}{deep learning}
    \acro{NLP}{natural language processing}
    \acro{MF}{Matrix Factorization}
    \acro{HR}{Hit Rate}
    \acro{DCG}{Discounted Cumulative Gain}
    \acro{NDCG}{Normalized Discounted Cumulative Gain}
    \acro{SVD}{Collaborative Filtering}
    \acro{RNN}{recurrent neural networks}
    \acro{GRU}{Gated Recurrent Unit}
    \acro{LSTM}{Long Short-Term Memory}
    \acro{CNN}{convolutional neural networks}
    \acro{GNN}{graph neural networks}
    \acro{ACT-R}{Adaptive Control of Thought—Rational}
    \acro{SR}{sequential recommendation}
    \acro{SVD}{Singular Value Decomposition}
    \acro{NBR}{Next Basket Recommendation}
    \acro{RepR}{Repetition Ratio}
    \acro{BPR}{Bayesian Personalized Ranking}
    \acro{PISA}{Psychology-Informed Session embedding using ACT-R}
    \acro{REACTA}{Recommendations from Embeddings with ACT-R and Audio features}
\end{acronym}
\begin{abstract}
On music streaming services, listening sessions are often composed of a balance of familiar and new tracks. Recently, sequential recommender systems have adopted cognitive-informed approaches, such as Adaptive Control of Thought—Rational (ACT-R), to successfully improve the prediction of the most relevant tracks for the next user session. However, one limitation of using a model inspired by human memory (or the past), is that it struggles to recommend new tracks that users have not previously listened to. To bridge this gap, here we propose a model that leverages audio information to predict in advance the ACT-R-like activation of new tracks and incorporates them into the recommendation scoring process. We demonstrate the empirical effectiveness of the proposed model using proprietary data,
which we publicly release along with the model’s source code to foster future research in this field.
\end{abstract}



\maketitle

\section{Introduction}
\label{sec_intro}
In recent years, while sequential recommendation systems~\cite{fang_tois20} have proven effective in music domain~\cite{tran_sigir23,moor_cikm23,pereira_recsys19,schedl2018current,bendada2023scalable}
, they often overlook, or inadequately model, \textit{repetitive} interaction patterns. This represents a significant limitation for music-focused applications~\cite{cheng_ijcai17, wang_inforet18, pereira_recsys19, hansen_recsys20, wang_multimedia21} where repeatedly listening to the same tracks over time is frequent~\cite{gabbolini_recsys21,sguerra2022discovery, sguerra2025uncertainty}.
Repeated exposure is not only typical, but instrumental in the music discovery process, shaping how users perceive and connect with individual tracks~\cite{sguerra2022discovery}.

One of the recent lines of research has focused on modeling repeat behavior in recommendation systems based on Anderson’s \ac{ACT-R} cognitive architecture~\cite{anderson_psyreview04, bothell20}. With applications spanning hashtag reuse, mobile app usage prediction, job recommendation, and modeling music genre preferences~\cite{lex2019impact, lex2020modeling, lacic2017beyond, lacic2019should, zhao2014context}. ACT-R is a well-established  
cognitive architecture and unified theory of cognition, designed to model the structure and processes of the human mind. It aims to explain human cognition in all its complexity through a fixed set of modules, particularly notable for its module that captures the dynamics of memory access. In the music domain, Reiter-Haas et al.~\cite{reiterhaas_recsys21} applied ACT-R's declarative memory module to predict music relistening behavior within user sessions, outperforming baselines that prioritized recency-based track selection. Moscati et al.~\cite{marta_recsys23} then pointed out that the model only recommended tracks having been previously listened by users. They expanded to integrate ACT-R with collaborative filtering approaches, such as \ac{BPR}~\cite{rendle_auai09}, to recommend both familiar and novel tracks. They first pre-trained a collaborative filtering model, then adjusted its recommendation scores using ACT-R during inference. More recently, Tran et al.~\cite{tran_recsys24} identified further shortcomings in these earlier efforts. They observed that ACT-R was applied exclusively at inference time, with no influence during model training. To address this, they introduced PISA (\underline{P}sychology-\underline{I}nformed \underline{S}ession embedding using \underline{A}CT-R), a model that integrates ACT-R activation into attention mechanisms during training to better capture both the \textit{dynamic} and \textit{repetitive} patterns in user behavior.

We contend that prior approaches suffer from a well-known limitation: since~\ac{ACT-R}'s declarative module models memory, it can only be applied to repeated tracks, leaving new tracks unaddressed. However, here we posit that unseen tracks should still retain some activation, not from memory, but based on a higher level representation of similar music. For instance, even if a user has never listened to a specific track by an artist, their past exposure to hits by similar artists should still influence their activation.

In this short paper, we aim to fill this gap. Our contributions are threefold: (1) We introduce a novel model that leverages audio features to predict \ac{ACT-R}-like activation, allowing the model to anticipate user engagement with both repeated and new tracks. (2) We demonstrate the suitability of our approach through extensive experiments on proprietary data from a global music streaming platform. (3) To promote transparency and foster further research, we release our source code and industrial-grade dataset, which includes longer user listening histories, reduced recommendation bias and is more aligned with users' intention by solely accounting for \textit{organic} (i.e. user-selected) interactions, and enriched audio embeddings w.r.t the one released in~\cite{tran_recsys24}.

\section{Preliminaries}
\subsection{Problem Formulation}
\label{prob_formulation}
Following the setting proposed in previous work ~\cite{hansen_recsys20, tran_recsys24}, we consider a set~$\mathcal{U}$ of users and a set~$\mathcal{V}$ of tracks in this paper. For each user $u \in \mathcal{U}$, we observe\footnote{For users with more than $L$ sessions, one may consider a subset of $L$ sessions, such as the most recent $L$.} an ordered sequence of $L \in \mathbb{N}^*$ past listening sessions, denoted by $S^{(u)} = (s^{(u)}_1, s^{(u)}_2, \dots, s^{(u)}_{L})$ where $s^{(u)}_l \in S^{(u)}$, with $l \in \{1,\dots,L\}$, corresponds to the $l$-th listening session of user $u$ and is represented as a set (unordered collection) of $K \in \mathbb{N}^*$ tracks\footnote{Consistent with \cite{hansen_recsys20, tran_recsys24}, we consider only the first $K$ tracks of each session.} that the user listened to during that session: $s^{(u)}_l = \{v^{(u)}_{l,1}, v^{(u)}_{l,2},...,v^{(u)}_{l,K}\}$, 
with $v^{(u)}_{l,k} \in \mathcal{V}, \forall k \in \{1,\dots,K\}$. The task is to predict: $s^{(u)}_{L+1} = \{v^{(u)}_{L+1, 1}, v^{(u)}_{L+1, 2},...,v^{(u)}_{L+1, K}\}$, i.e., the set of $K$ tracks that $u$ will interact with in their next session $s^{(u)}_{L+1}$, based on $S^{(u)}$.

In addition, each track in the set $\mathcal{V}$ is associated with two pre-trained embedding matrices: $\mathbf{M} \in \mathbb{R}^{|\mathcal{V}| \times d}$ and $\mathbf{A} \in \mathbb{R}^{|\mathcal{V}| \times d'}$. $\mathbf{M}$ is calculated from the co-occurrences of tracks in diverse music collections (e.g., playlists) using \ac{SVD}~\cite{koren_computer09}, while $\mathbf{A}$ consists of audio-based embeddings~\cite{meseguer_icassp24}, respectively. Each row of $\mathbf{M}$ (resp. $\mathbf{A}$) provides an embedding vector $\mathbf{m}_v \in \mathbb{R}^d$ (resp. $\mathbf{a}_v \in \mathbb{R}^{d'}$) representing a track $v \in \mathcal{V}$, with $d, d' \in \mathbb{N}^*$ denoting the respective embedding dimensions.

\subsection{ACT-R Framework}
  
The ACT-R declarative module comprises a set of activation functions that simulate how the human mind retrieves stored information, and it has shown notable success in modeling repetitive behaviors~\cite{lex2019impact, szpunar2004liking, peretz1998exposure, sguerra2023ex2vec}. Specifically, to estimate how easily a user $u \in \mathcal{U}$ can retrieve a track $v \in \mathcal{V}$ from memory, the module computes a sum of component values, each capturing a distinct cognitive factor influencing memory access~\cite{bothell20}:
\subsubsection{Base-level component}
$\text{BL}^{(u)}_v$ captures the principle that information accessed more frequently or more recently is more easily retrieved from memory~\cite{reiterhaas_recsys21, marta_recsys23, tran_recsys24}. We set:
    \begin{equation}
\label{bll_eq}
    \text{BL}^{(u)}_v = \text{softmax}_{s_l^{(u)}}\Big(\sum_{k}(t_{s_l^{(u)}} - t^{(uv)}_{k})^{-\alpha}\Big).
\end{equation}
where $t_{s_l^{(u)}}$ represents the start time of session $s_l^{(u)}$, and $t^{(uv)}_{k}$ denotes the time of the $k$-th instance in which user $u$ listened to track $v$, with $t^{(uv)}_{k} < t_{s_l^{(u)}}$. The parameter $\alpha \in \mathbb{R}^+$ acts as a time decay factor, capturing the effect of memory decay for past listens. A softmax operation is applied to normalize the resulting scores across all tracks in the session, ensuring that $\sum_{v \in s_l^{(u)}} \text{BL}^{(u)}_v = 1$.

\subsubsection{Spreading component}
$\text{SPR}^{(u)}_v$ spreads activation across items based on contextual information, specifically, session co-occurrence patterns. It is grounded in the idea that if a track $v$ is frequently accompanied by certain tracks in a given context (past sessions), then the presence of those tracks in the most recent session will boost the memory activation of $v$ during the current session. In the same fashion as \cite{tran_recsys24}, for each track $v \in s_{l}^{(u)}$, we define
    \begin{equation}
    \text{SPR}^{(u)}_v = \sum_{v' \in s_{l-1}^{(u)}} \textbf{C}_{v'v}.
    \label{spr}
\end{equation}
We use the track correlation matrix $\mathbf{C} = \mathbf{D}^{-\frac{1}{2}} \mathbf{F} \mathbf{D}^{-\frac{1}{2}}$, where $\mathbf{D}$ is a diagonal matrix with entries $\mathbf{D}_{ii} = \sum_j \mathbf{F}_{ij}$ for all $i$, and $\mathbf{D}_{ij} = 0$ for all $i \neq j$~\cite{le_ijcai19}. The matrix $\mathbf{F} \in \mathbb{R}^{|\mathcal{V}| \times |\mathcal{V}|}$ denotes the track co-occurrence matrix, where $\mathbf{F}_{ij}$ captures the number of times track $j$ appeared in the session immediately preceding a session containing track $i$.

\subsubsection{Partial matching component}
$\text{P}^{(u)}_v$ enhances memory activation by accounting for  similarity between tracks. For example, if track $v$ is a jazz song, the presence of a musically similar jazz track $v'$ in the most recent session can boost the activation of $v$. This similarity-based activation is computed using the dot products of \ac{SVD}-based embedding vectors for each track $v$ in the session $s_l^{(u)}$:
\begin{equation}
    \text{P}^{(u)}_v = \sum_{v' \in s_{l-1}^{(u)}} \mathbf{m}^{\intercal}_v \mathbf{m}_{v'}.
\end{equation}

\subsection{Psychology-Informed Session embedding using ACT-R (PISA)}
Tran et al. \cite{tran_recsys24} introduced PISA, a Transformer-based method designed for repeat-aware and sequential listening session recommendation. PISA utilizes attention mechanisms inspired by ACT-R components to capture embedding representations of sessions and users, effectively modeling both sequential and repetitive patterns in historical listening behavior.

\begin{figure*}[ht]
  \centering
\includegraphics[width=0.86\textwidth]
{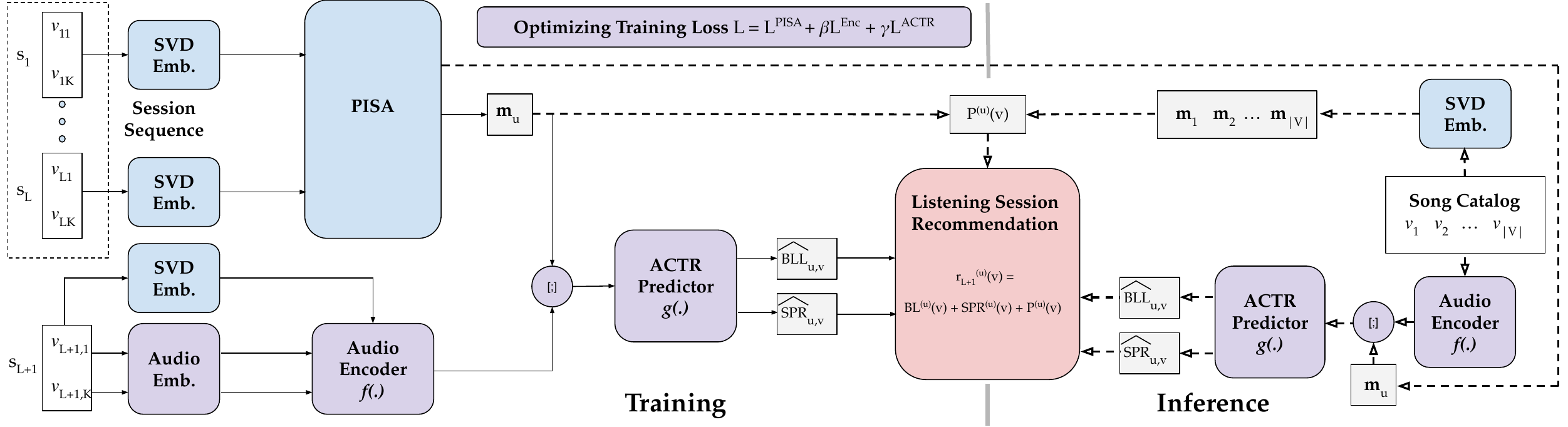}
  \caption{Architecture of REACTA model (dashed arrows are for inference time).}
  \label{fig:au2actr}
\end{figure*}

\subsubsection{Session Embedding}
Given the track embedding matrix $\mathbf{M} \in \mathbb{R}^{|\mathcal{V}| \times d}$, PISA learns embedding representations for session $s_l^{(u)}$ of some user $u \in \mathcal{U}$, denoted as $\mathbf{m}_{s_l^{(u)}} \in \mathbb{R}^d$, using attention weights guided by ACT-R components as follows:

\begin{equation}
    \mathbf{m}_{s_l^{(u)}} = \sum_{v \in s_l^{(u)}} w_v \mathbf{m}_{v}
\end{equation}

The terms $w_v \geq 0$, with $\sum_{v \in s_l^{(u)}} w_v = 1$, are ACT-R-informed attention weights associated with each track in the session, with:

\begin{equation}
    w_v = w_{\text{BL}}\text{BL}^{(u)}_{v} + w_{\text{SPR}} \text{SPR}^{(u)}_{v} + w_{\text{P}} \text{P}^{(u)}_{v}
\end{equation}


\subsubsection{User Embedding}
PISA integrates both short-term and long-term preferences to compute the final user representation:
\begin{equation}
    \mathbf{m}_u = \beta \mathbf{m}^{\text{short}}_u + (1 - \beta) \mathbf{m}^{\text{long}}_u
\end{equation}
where the parameter $\beta \in [0, 1]$ is learned using a one-layer feedforward neural network applied to the concatenation $[\mathbf{m}^{\text{short}}_u; \mathbf{m}^{\text{long}}_u]$. The vector $\mathbf{m}^{\text{long}}_u \in \mathbb{R}^d$, capturing the user's ``long-term'' preferences, independent of contextual factors; while the vector $\mathbf{m}^{\text{short}}_u \in \mathbb{R}^d$, reflecting the influence of recent listening sessions on the user’s ``short-term'' preferences and recommendation perception. This component is modeled dynamically using a Transformer~\cite{vaswani_nips17} applied over sequences of past sessions.

\section{Proposition}
\label{section_proposition}
As discussed in Section~\ref{sec_intro}, previous \ac{ACT-R}-based methods face a key limitation: activation is restricted to previously seen tracks, meaning these methods cannot handle new ones, as the declarative memory module in \ac{ACT-R} models only memory recall. In particular, the base-level component $\text{BL}^{(u)}_v$ in Equation~\ref{bll_eq} and the spreading activation component $\text{SPR}^{(u)}_v$ in Equation~\ref{spr} are computed solely for tracks present in a user's listening history, rendering them undefined for unseen tracks.

To overcome this limitation, we propose REACTA (\underline{R}ecommenda-tions from \underline{E}mbeddings with \underline{ACT}-R and \underline{A}udio features), by building on top of PISA with two additional components: an audio encoder and a predictor for ACT-R-like activation. The overall architecture is shown in Figure~\ref{fig:au2actr}.

\subsection{Audio Encoder}
We employ a two-layer feedforward neural network $f$ to project the audio embedding $\mathbf{a}_v \in \mathbb{R}^{d'}$ of each track $v$ into a vector $f(\mathbf{a}_v) \in \mathbb{R}^d$ (recall that $d$ is the dimension of \ac{SVD}-based embeddings). We also introduce a constraint that promotes the proximity of the encoded vector to the \ac{SVD}-based embedding $\mathbf{m}_v$ of the corresponding track, while simultaneously distancing it from 
a $\mathbf{m}_{v'}$ negative sample (i.e., a different track). 

\subsection{ACT-R-like Activation Predictor}
Inspired by the work of Briand et al.~\cite{briand2024let},
which predicts \ac{SVD}-based embeddings of newly released tracks using metadata, we estimate ACT-R-like activations of a pair user $u$  and track $v$ based on audio features. Specifically, we concatenate the audio encoder output for each track in the session $s^{(u)}_{l+1}$ with the embedding of the previous session $s^{(u)}_{l}$, computed by the PISA component using only the first $l$ sessions, forming the input $[f(\mathbf{a}_v); \mathbf{m}_{u,l}]$. A two-layer feedforward neural network $g$ is used to map this input into the ACT-R weight space, which consists of the concatenated base-level component $\text{BL}^{(u)}_v$ and the spreading component $\text{SPR}^{(u)}_v$. These predicted weights are then used at inference time to compute the final recommendation scores. It’s worth noting that the partial matching component is excluded here, as it is accounted for by another term in the scoring function, which will be explained in Section~\ref{reco_scoring}.

\begin{table*}[ht]
\caption{Listening session recommendation results. All metrics should be maximized, except MR (minimized), RepBias (close to 0). Bold and underlined numbers correspond to the best and second-best performance for each metric, respectively.}
  \label{acc_results}
  \resizebox{\textwidth}{!}{
  \begin{tabular}{c|c||cc|cc|cc|cc}
        \toprule
        \multirow{2}{*}{\textbf{Dataset}} &  \multirow{2}{*}{\textbf{Model}} & \multicolumn{2}{c}{\textbf{Global Metrics}} & \multicolumn{2}{c}{\textbf{Repetition-Focused Metrics}} & \multicolumn{2}{c}{\textbf{Exploration-Focused Metrics}} & \multicolumn{2}{c}{\textbf{Beyond-Accuracy Metrics}} \\
        \cline{3-10}
         &  & {NDCG (in \%)} & {Recall (in \%)} & {$\text{NDCG}^{\text{Rep}} \text{ (in \%)}$} & {$\text{Recall}^{\text{Rep}}\text{ (in \%)}$} & {$\text{NDCG}^{\text{Exp}}\text{ (in \%)}$} & {$\text{Recall}^{\text{Exp}}\text{ (in \%)}$} & RepBias & PopBias \\
        \midrule
        \midrule
        \multirow{6}{*}{\shortstack{\textbf{Proprietary Dataset} \\ \\ RepRatio-GT = 83.79\%}} & {ACT-R-Repeat} & {\textbf{10.74 $\pm$ 0.17}} & {\textbf{10.34 $\pm$ 0.18}} & {\textbf{11.70 $\pm$ 0.17}} & {\textbf{11.90 $\pm$ 0.18}} & {0.00 $\pm$ 0.00} & {0.00 $\pm$ 0.00} & {16.21 $\pm$ 0.00} & {28.05 $\pm$ 0.15} \\
        {} &  {ACT-R-BPR} & {7.41 $\pm$ 0.16} & {6.92 $\pm$ 0.16} & {7.26 $\pm$ 0.15} & {7.07 $\pm$ 0.15} & {1.78 $\pm$ 0.08} & {2.59 $\pm$ 0.10} & {-21.28 $\pm$ 0.23} & {\textbf{6.46 $\pm$ 0.04}} \\      
        {} &  {PISA-U} & {8.34 $\pm$ 0.12} & {7.72 $\pm$ 0.11} & {8.39 $\pm$ 0.13} & {8.26 $\pm$ 0.14} & {2.01 $\pm$ 0.06} & {2.89 $\pm$ 0.06} & {1.87 $\pm$ 0.11} & {27.88 $\pm$ 0.17} \\
        {} &  {PISA-P} & {8.88 $\pm$ 0.11} & {8.19 $\pm$ 0.12} & {8.86 $\pm$ 0.13} & {8.68 $\pm$ 0.13} & {2.36 $\pm$ 0.05} & {3.34 $\pm$ 0.08} & {1.48 $\pm$ 0.09} & {25.54 $\pm$ 0.16} \\
        \cline{2-10}  
        {} & {REACTA-U (ours)} & {8.56 $\pm$ 0.12} & {7.87 $\pm$ 0.11} & {8.45 $\pm$ 0.13} & {8.26 $\pm$ 0.14} & {2.65 $\pm$ 0.06} & {3.71 $\pm$ 0.03} & {0.33 $\pm$ 0.12} & {27.20 $\pm$ 0.15} \\
        {} &  {REACTA-P (ours)} & \underline{\textit{9.35 $\pm$ 0.14}} & \underline{\textit{8.57 $\pm$ 0.14}} & {\underline{\textit{9.10 $\pm$ 0.16}}} & {\underline{\textit{8.89 $\pm$ 0.16}}} & {\textbf{3.30 $\pm$ 0.08}} & {\textbf{4.52 $\pm$ 0.11}} & \textbf{{0.09 $\pm$ 0.08}} & \underline{\textit{23.96 $\pm$ 0.15}} \\
        \bottomrule
      \end{tabular}
    }
\end{table*}

\subsection{Listening Session Recommendation}
\label{reco_scoring}
To predict the set of tracks that user $u$ is likely to listen to in the next session, following $S^{(u)}$,
we adopt a two-stage approach to obtain the relevance score of each track $v \in \mathcal{V}$ for a user $u \in \mathcal{U}$. In the first stage, we estimate ACT-R-like weights ($\text{BL}^{(u)}_v$ and $\text{SPR}^{(u)}_v$) from audio embeddings for all new tracks $v$ that user $u$ has not previously interacted with, allowing us to obtain complete base-level and spreading components for every item in the catalog, based on a higher level representation of music from the audio embeddings. In the second stage, we compute the remaining partial matching component $P^{(u)}_v$ using the dot product: $P^{(u)}_v = \mathbf{m}_{u}^{\intercal} \mathbf{m}_v$. The final relevance score for each track $v \in \mathcal{V}$ is then defined as the sum of these components, forming the complete ACT-R activation:
\begin{equation}
    r^{(u)}_{L+1}(v) = \text{BL}^{(u)}_v + \text{SPR}^{(u)}_v + P^{(u)}_v
\end{equation}

\subsection{Training Procedure}
\label{training_secsion}
We use a dataset $\mathcal{S}$ consisting of session sequences to optimize $\Theta$, the full set of model parameters. For each sequence $S^{(u)} = (s^{(u)}_1, s^{(u)}_2, \dots, s^{(u)}_{L}) \in \mathcal{S}$, we generate sub-sequences containing only the first $l$ sessions, where $l \in \{1, \dots, L-1\}$.
When recommending a set of $K$ tracks to extend this truncated sequence, the model is expected to assign high relevance scores to the tracks in $s^{(u)}_{l+1}$, i.e., the \textit{ground truth} next-session tracks while assigning lower scores to those in $o^{(u)}_{l+1}$, a randomly sampled set of $K$ \textit{negative} examples drawn from $\mathcal{V} \setminus s^{(u)}_{l+1}$. To this end, we adopt a multi-task training approach and optimize $\Theta$ via gradient descent by minimizing the loss function:

\begin{equation}
\label{loss}
    \mathcal{L}(\Theta) = \mathcal{L}^{\text{PISA}}(\Theta) + 
\beta \mathcal{L}^{\text{Enc}}(\Theta) + \gamma \mathcal{L}^{\text{ACTR}}(\Theta)
\end{equation}
\begin{align*}
    \mathcal{L}^{\text{PISA}}(\Theta) &= \lambda \sum_{S^{(u)} \in \mathcal{S}} \sum^{L-1}_{l=1} \sum_{v \in s^{(u)}_{l+1}, v' \in o^{(u)}_{l+1}} \ln\bigl(1 + e^{-(\mathbf{m}^{\intercal}_{u,l}\mathbf{m}_v - \mathbf{m}^{\intercal}_{u,l}\mathbf{m}_{v'})}\bigr) \nonumber \\
    &+ (1 - \lambda) \sum_{S^{(u)} \in \mathcal{S}} \sum^{L-1}_{l=1} \bigl(1 - \mathbf{m}^{\intercal}_{u,l}\mathbf{m}_{s^{(u)}_{l+1}}\bigr),\nonumber\\
\end{align*}
\begin{align*}
    \mathcal{L}^{\text{Enc}}(\Theta) &= \sum_{S^{(u)} \in \mathcal{S}} \sum^{L-1}_{l=1} \sum_{v \in s^{(u)}_{l+1}, v' \in o^{(u)}_{l+1}} \ln\bigl(1 + e^{-(f(\mathbf{a}_{v})^{\intercal}\mathbf{m}_v - f(\mathbf{a}_{v})^{\intercal}\mathbf{m}_{v'})}\bigr),\nonumber\\
    \mathcal{L}^{\text{ACTR}}(\Theta) &= \sum_{S^{(u)} \in \mathcal{S}} \sum^{L-1}_{l=1} \sum_{v \in s^{(u)}_{l+1}} ||g([f(\mathbf{a}_v); \mathbf{m}_{u,l}]) - [\text{BL}_v^{(u)}; \text{SPR}_v^{(u)}]||^2_2 \nonumber
\end{align*}
where $\lambda$, $\beta$ and $\gamma$ are hyper parameters.
    

\section{Experimental Analysis}
\label{experiment}
\subsection{Dataset}
We conduct an extensive evaluation of next session recommendation on a large-scale proprietary dataset from the music domain. This dataset comprises nearly 900 million time-stamped listening events—collected over the course of one year from more than 4 million users of the music streaming service Deezer. 
Only user-selected interactions are included for two reasons: first, to mitigate biases introduced by recommendation algorithms; and second, because we posit that such interactions, which require active engagement, better reflect true user intent. In contrast, interactions with algorithmic suggestions may involve more passive engagement, making intent less reliable.
A listening event is defined as a user streaming a track for at least 30 seconds, a standard threshold widely adopted in the industry for remuneration purposes. The dataset contains 50,000 tracks, representing the most popular content on the platform during the year 2023. In addition to interaction logs, we also provide pre-trained audio embeddings \cite{meseguer_icassp24} and \ac{SVD}-based embeddings~\cite{briand2024let} for each track in the collection. The dataset is publicly available on our GitHub repository\footnote{\label{sourcecode}https://github.com/deezer/recsys25-reacta}.
\label{datasets}
\subsection{Task and Evaluation Metrics}
\subsubsection{Task}
We use the last 20 sequences of each user, randomly splitting them into 10 for validation and 10 for testing. Within each sequence, we observe the first \( L = 30 \) sessions, while the \( 31^{\text{st}} \) session is masked and used as the prediction target. We assess the ability of our proposed model and baseline methods to accurately retrieve the \( K = 10 \) tracks from the masked session, ranked by predicted relevance scores, based on the preceding sessions. 

\subsubsection{Evaluation}
Following prior work~\cite{tran_recsys24}, we evaluate each model using eight metrics. Six focus on accuracy: global metrics (Recall, \ac{NDCG}), repetition-focused metrics ($\text{Recall}^{\text{Rep}}$, $\text{NDCG}^{\text{Rep}}$), and exploration-focused metrics ($\text{Recall}^{\text{Exp}}$, $\text{NDCG}^{\text{Exp}}$). The remaining two metrics capture beyond-accuracy aspects of recommendation quality: RepBias measures the difference in repetition rate between the recommended and ground truth sessions (RepRatio-GT), while PopBias quantifies the intra-session median rank of the tracks in the recommended session, reflecting popularity bias.

\subsection{Models}
\subsubsection{Two variants of our proposition}
We extend two variants of PISA from \cite{tran_recsys24}, both built upon the architecture described in Section~\ref{section_proposition}, but differing in their negative sampling strategies used during training to evaluate the loss in Equation~\eqref{loss}. The first variant, denoted REACTA-U, uniformly samples 10 tracks for each negative set $o^{(u)}_{l+1}$ from the set of unlistened tracks $\mathcal{V} \setminus s^{(u)}_{l+1}$. The second variant, REACTA-P, uses a popularity-based negative sampling strategy,
where more popular tracks are more likely to be selected as negative samples.

\subsubsection{Baselines}
We compare REACTA against four baseline models representing all existing ACT-R-based approaches in the music domain. ACT-R-Repeat, proposed by Reiter-Haas et al.~\cite{reiterhaas_recsys21}, recommends only repeated tracks. ACT-R-BPR, introduced by Moscati et al.~\cite{marta_recsys23}, extends ACT-R-Repeat by incorporating \ac{BPR}~\cite{rendle_auai09} to recommend both repeated and novel tracks. The remaining baselines, PISA-U and PISA-P, are two variants developed by Tran et al.~\cite{tran_recsys24}.

\subsubsection{Implementation Details}
We train REACTA-U, REACTA-P and other baselines for a maximum of 100 epochs using the Adam optimizer~\cite{kingma_iclr15} and batch sizes of 512. We set embedding dimension $d =$ 128, $\alpha = 1/2$ for the BL module of all ACT-R models. We also set sequence's length $L=30$, number of blocks $B = 2$ and number of heads $H = 2$ for Transformer-based models. Other hyperparameters were tuned via grid search on the validation set. Most notably, we test learning rates values in $\{{0.0002, 0.0005, 0.00075, 0.001}\}$, $\lambda$ values in $\{0.0, 0.3, 0.5, 0.8, 0.9, 1.0\}$, and $\beta$, $\gamma$ values in $\{0.2, 0.4, 0.6, 0.8, $ $1.0\}$.



\subsection{Results and Discussion}
\label{results}

Table~\ref{acc_results} summarizes all test results, averaged across five runs along with their standard deviations. Overall, REACTA demonstrates competitive performance, particularly strong on exploration-focused metrics and effectively aligning recommendations with user behavior in terms of repetition and exploration. 

\subsubsection{REACTA vs Other ACT-R Methods}

REACTA-P consistently ranks among the top performers across four global accuracy metrics. While baselines like ACT-R-Repeat excel at recommending familiar tracks, they entirely neglect novel content. ACT-R-BPR introduces exploration via collaborative signals but sacrifices repetition accuracy. PISA-U (and P) strike a better balance, improving on both fronts compared to ACT-R-BPR. REACTA-U (and P) match PISA’s performance on repetition metrics but significantly outperform all baselines in recommending unheard tracks. Notably, REACTA-P achieves a top $\text{NDCG}^{\text{Exp}}$ of 3.30\%, highlighting its strength in exploration—a key factor for music discovery. These results demonstrate that combining session-based ACT-R activation with predicted activations for unseen tracks enhances both repetitive and exploratory recommendation quality.


\subsubsection{On Repetition and Popularity Biases}
Beyond performance, balancing familiar and novel tracks in each session is key for effective personalization. We note that the ground truth average proportion of repeated tracks in test sessions to retrieve, i.e., RepRatio-GT, is relatively high (83.79\%). The RepBias metric confirms that ACT-R-Repeat is, as expected, biased toward repetition, while ACT-R-BPR leans toward exploration. In comparison, PISA-U (and P) better align with user consumption patterns. Notably, REACTA-U (and P) achieve the best balance, closely matching ground-truth repetition ratios, with RepBias as low as 0.09\%.

Besides, popularity-based negative sampling helps reduce models' susceptibility to popularity bias, as seen in both PISA and REACTA. Still, ACT-R-BPR remains a strong baseline, outperforming other ACT-R-based methods in this aspect with the lowest PopBias score of 6.46.

\section{Conclusion}
We introduced REACTA, a model that estimates ACT-R-like activation for new tracks using audio similarity, integrating this signal into recommendation scoring. This addresses a key limitation of memory-based methods, which compute activation only for previously heard tracks. Experiments show that REACTA performs well in warm-start scenarios. Moreover, it shows promise for cold-start settings by substituting audio encoder's representations for missing \ac{SVD} embeddings at inference.

The main limitation of our approach is the computational cost of calculating ACT-R activation over a large catalog. We plan to address this in future work using approximate activation methods~\cite{petrov2006computationally}.

Finally, given the interpretability of ACT-R activation scores, REACTA could be extended to let users express preferences for exploration or repetition across contexts. The model could then re-weight or truncate the item activations to emphasize either novel or familiar content in session representations, enabling user control over the exploration-repetition balance.

\bibliographystyle{ACM-Reference-Format}

\end{document}